\documentclass[a4paper]{jpconf}
\usepackage[utf8]{inputenc}
\usepackage{placeins}
\usepackage{geometry}
 \usepackage{graphicx}
 \usepackage{subfig}
  \usepackage{amsmath} 
  \graphicspath{{images/}}
  \usepackage[usenames, dvipsnames]{color}
\usepackage{wrapfig}
\usepackage{boxedminipage}
\usepackage{multirow}

\usepackage{fullpage}
\usepackage{amsmath}
\usepackage{amssymb}
\usepackage{graphicx}
\usepackage{mathrsfs}
\usepackage{wrapfig}
\usepackage{boxedminipage}
\usepackage{epsfig}
\usepackage{setspace}
\usepackage{float}
\usepackage{hyperref}
\usepackage{enumerate}
\usepackage{cite}
\usepackage[font=footnotesize,labelfont=rm]{caption}

\usepackage[numbers,sort&compress]{natbib}

\def\be{\begin{equation}}
\def\ee{\end{equation}}
\def\bea{\begin{eqnarray}}
\def\eea{\end{eqnarray}}

\numberwithin{equation}{section}

 \newcommand{\RN}[1]{%
   \textup{\uppercase\expandafter{\romannumeral#1}}%
 }

\usepackage{graphicx}
\usepackage{iopams}
\begin{document}
\title{Topology of Hawking-Page transition in Born-Infeld AdS black holes}

\author{Pavan Kumar Yerra$^{1,2}$, Chandrasekhar Bhamidipati$^3$ and
	Sudipta Mukherji$^{1,2}$}

\address{$^1$ Institute of Physics, Sachivalaya Marg, Bhubaneswar, Odisha, 751005, India }
\address{$^2$ Homi Bhabha National Institute, Training School Complex, Anushakti Nagar, Mumbai, 400085, India}
\address{$^3$ School of Basic Sciences,
Indian Institute of Technology Bhubaneswar, Bhubaneswar, Odisha, 752050, India}

\ead{pk11@iitbbs.ac.in,chandrasekhar@iitbbs.ac.in,mukherji@iopb.res.in}

\begin{abstract}
Black holes in anti de Sitter (AdS) spacetimes undergo phase transitions which typically lead to the existence of critical points, that can be classified using topological techniques. Availing the Bragg-Williams construction which provides an off-shell free energy formalism, we compute the topological charge of the Hawking-Page (HP) transition for Einstein-Born-Infeld black holes in anti de Sitter spacetime and match the result with the confinement-deconfinement transition in the dual gauge theory, which turn out to be in perfect agreement. 

\end{abstract}

\section{Introduction}

Thermodynamics of Black holes and their phase transitions, particularly in anti de Sitter space times (AdS) have continued to provide interesting insights~\cite{Bekenstein:1973ur,Bardeen:1973gs,Hawking:1975vcx,Chamblin:1999tk,Caldarelli:1999xj,Kastor:2009wy,Cvetic:2010jb,Dolan:2011xt,Karch:2015rpa,Kubiznak:2012wp,Kubiznak:2016qmn,Banerjee:2011cz,Lu:2010xt}, with a special place for the Hawking-Page transition~\cite{Hawking:1982dh}, which can be interpreted as a confinement-deconfinement transition in the dual gauge/gravity setting~\cite{,Maldacena:1997re,Gubser:1998bc,Witten:1998qj}. Mean field approximation turns out to be a reliable approach to study phase transitions, not just in black holes, but for general thermodynamic systems~\cite{cha95,bw1, bw2,kubo,Banerjee:2010ve}. A particularly elegant approach is the one by Bragg-Williams, where one starts by constructing an off shell free energy given in terms of an order parameter. Minimising this free energy then gives the required phases of the thermodynamic system. This approach has been applied successfully to black holes in AdS as well~\cite{Banerjee:2010ve,Chandrasekhar:2012vh}.\\

\noindent
Novel topological techniques have be pioneered to classify critical points of black holes\footnote{see~\cite{Cunha:2017qtt,Cunha:2020azh,Guo:2020qwk,Wei:2020rbh,Junior:2021svb,Wu:2023eml} for other developments  which were initiated earlier from the study of light-rings}~\cite{Duan:1984ws,Duan:2018rbd,Wei:2021vdx}. The method and the results have now been generalised to several interesting situations~\cite{Yerra:2022alz,Ahmed:2022kyv,Yerra:2022eov,Wei:2022dzw,Barzi:2023msl,Gogoi:2023qku,Alipour:2023uzo,Gogoi:2023xzy,Wu:2022whe,Fang:2022rsb,Fan:2022bsq,Liu:2022aqt,Bai:2022klw,Zhang:2023tlq,Hung:2023ggz,Wang:2023qxw,Chen:2023pqk,Sadeghi:2023dsg,Shahzad:2023cis}. It is now understood that topological charges can be assigned not just to first and second order phase transition points, but to the black hole solutions themselves. In particular, it has now been possible to compute the topological charge of the confinement-deconfinement transition using an effective potential in the dual gauge theory and match the result with the charge of Hawking-Page transition point in the bulk~\cite{Yerra:2022coh}. For the case of charged black hole in AdS, it is found that the two topological charges are same, even though the order parameters in the bulk and boundary are quite different, namely, horizon radius $r_+$ for the Hawking-Page transition point, and a charge parameter $Q$ for the confinement-deconfinement transition in gauge theory. There are further checks possible, knowing that the saddle points of the off shell Bragg-Williams free energy give the equilibrium phases of the thermodynamic system~\cite{cha95,Banerjee:2010ng,Banerjee:2010ve,nayak2008bragg}. Using the inputs from~\cite{Wei:2021vdx}, one can assign topological charges to stable and unstable phases of the system, which turn out to have opposite values~\cite{Wei:2022dzw}, both in the bulk and in the boundary~\cite{Yerra:2022coh,Yerra:2023hui}. The aim of this short article is to show that the computation of topological charges of Hawking-Page transition point and equilibrium phases of black holes done in~\cite{Yerra:2022coh}, can be extended to the case of Born-Infeld AdS black holes~\cite{Dey:2004yt} in the grand canonical ensemble. \\

\noindent
Rest of the article is structured as follows. In section-(\ref{2}), we present the Bragg-Williams off-shell free energy for Born-Infeld AdS black holes and locate the Hawking-Page transition point. Subsection-(\ref{2.1}) contains the computation of topological charge of the HP point. Section-(\ref{3}) contains the effective potential of the boundary gauge theory and the computation of topological charge of confinement-deconfinement transition point. We conclude in section-(\ref{4}) with some open questions.

\section{Bulk} \label{2}
 We start with the Bragg-Williams off-shell free energy $f$ of Born-Infeld AdS black holes in grand canonical ensemble (i.e., fixed potential $\mu$) in $(n+1)$-dimensional spacetime, given by~\cite{Dey:2004yt,Banerjee:2010ng}:
\begin{equation}\label{eq: BWf }
f(r_+,  T, \mu) = M - TS-\mu Q, 
\end{equation}
where, the mass $M$, entropy $S$,  potential $\mu$ and charge $Q$ of the black hole  are given in terms of the horizon radius $r_+$ as:
\begin{eqnarray}\label{eq:thermodynamics}
M  & = &  \omega_{n-1}(n-1)  \bigg\{r_+^{n-2}+\frac{r_+^n}{l^2}+\frac{4 \beta ^2 r_+^n }{n (n-1)}\left(1-\sqrt{1+\frac{(n-2) (n-1) q^2}{2 \beta ^2 r_+^{2 n-2}}}\right)   \nonumber \\ 
&& \, \, \, +\frac{2 (n-1) q^2}{n r_+^{n-2}} \, _2F_1\big[\frac{n-2}{2 (n-1)},\frac{1}{2},\frac{3 n-4}{2 (n-1)},-\frac{(n-1) (n-2) q^2}{2 \beta ^2 r_+^{2 n-2}}\big] \bigg\},  \nonumber \\
 S &=& 4\pi \omega_{n-1} r_+^{n-1},   \nonumber \\
\mu & = & \frac{q}{c r_+^{n-2}} \, _2F_1\big[\frac{n-2}{2 (n-1)},\frac{1}{2},\frac{3 n-4}{2 (n-1)},-\frac{(n-1) (n-2) q^2}{2 \beta ^2 r_+^{2 n-2}}\big],  \nonumber \\
 Q & = & 2 \sqrt{2 (n-1) (n-2)} \, \omega_{n-1} \, q ,
\end{eqnarray}
with $\omega_{n-1}$ being the volume of a unit $(n-1)$ sphere, $c=\sqrt{\frac{2(n-2)}{n-1}}$, $l$ is the AdS length, $\beta$ is the Born-Infeld coupling, $_2F_1$ is the hypergeometric function, and $q$ is an integration constant related to the charge $Q$.
\begin{figure}[h!]
	
	{\centering
		
		\subfloat[]{\includegraphics[width=2.8in]{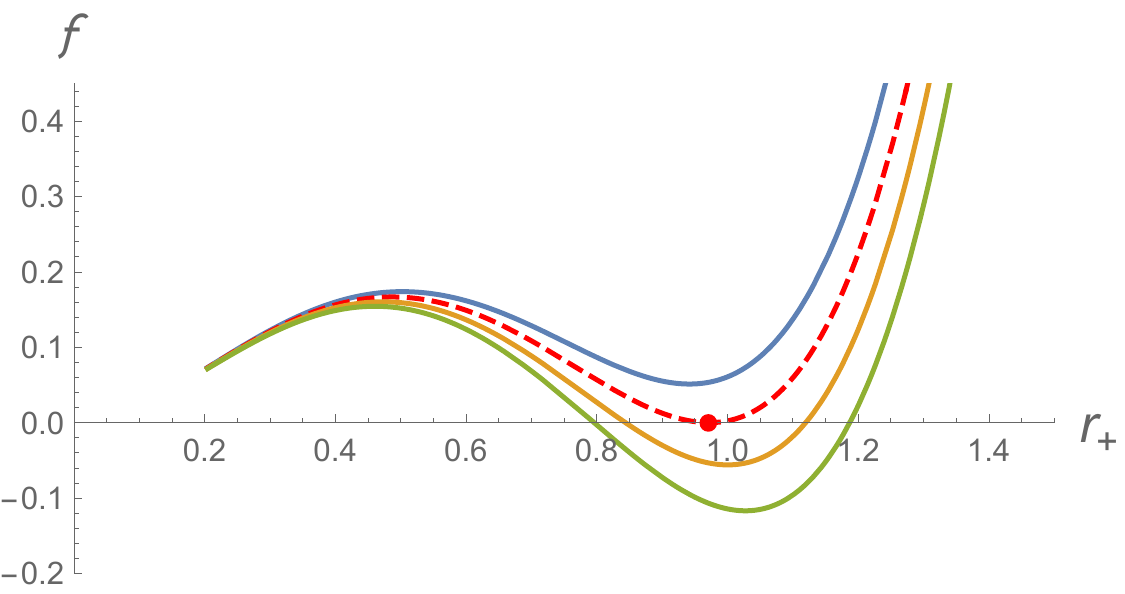}\label{fig:bulk_f}}\hspace{0.5cm}	
		\subfloat[]{\includegraphics[width=2.6in]{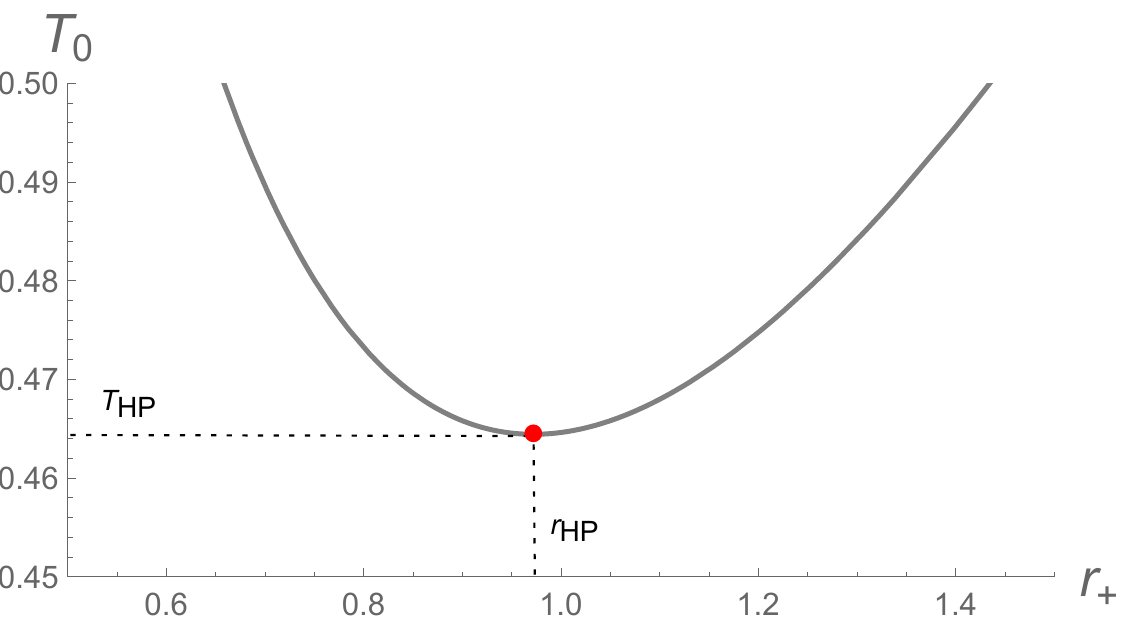}\label{fig:bulk_T0}}	
		
		\caption{\footnotesize  (a)  Bragg-Williams free energy $f$ is plotted against $r_+$ (using $x$ as a parameter)  at different temperatures $T$ for a fixed $\mu$. Temperature of the curves increases from top to bottom. Hawking-Page transition happens for  the temperature $T_{\rm HP}$ (dashed red curve) at  $r_{\rm HP}$ (red dot).  (b) The  curve $T_{\rm 0}$, as a function of $r_+$ (using $x$ as a parameter), shows the HP transition point at its minima (red dot).  Here, $(T_{\rm HP}, r_{\rm HP}, x_{\rm HP})= (0.464, 0.972, 0.243)$. Plots are displayed for $n=4, \mu=0.2, \omega_3 =l=\beta=1$.}
	}
	
\end{figure}
\\ \noindent
Treating horizon radius $r_+$ as order parameter, $T$ and $\mu$ as the external parameters, the behaviour of the free energy $f$ (plotted in the Fig.~\ref{fig:bulk_f}), shows the Hawking-Page (HP) transition, which happens when: 
\begin{equation}\label{eq:hp_condition}
f = 0, \, \text{and}, \, \frac{\partial f}{\partial r_+} = 0.
\end{equation} 
\noindent
Alternatively (which will be useful for later purpose), one can also find the Hawking-Page transition point using the first condition of eqn.~\ref{eq:hp_condition}, which gives, 
\begin{equation}\label{eq:bulk_T0}
f=0 \implies T=T_{\rm 0}^{\phantom{0}}(r_+, \mu).
\end{equation}
\noindent
The Hawking-Page transition point can be located at the  minima of the curve $T_{\rm 0}^{\phantom{0}}$, as shown in the Fig.~\ref{fig:bulk_T0}.
 We note here that,  due to the non-linear relation between $\mu$ and $q$ as in eqn.~\ref{eq:thermodynamics}, we study the free energy $f$ by defining a parameter $x$ as~\cite{Banerjee:2010ng}:
\begin{equation}\label{eq:x}
	x = \frac{q}{r_+^{n-1}}.
\end{equation}
The horizon radius, $r_+$ can now be rewritten as
\begin{equation}
	r_+ =  \frac{\mu c} {x \, _2F_1\big[\frac{n-2}{2 (n-1)},\frac{1}{2},\frac{3 n-4}{2 (n-1)},-\frac{(n-1) (n-2) x^2}{2 \beta ^2 }\big]}. 
\end{equation}
The HP point can be defined now using:
\begin{equation}
f = 0, \, \text{and}, \, \frac{\partial f}{\partial x} = 0,
\end{equation} 
\noindent
or, from the minima of the curve $T_0(x, \mu)$.

\subsection{ Topological charge of Hawking-Page transition point} \label{2.1}
\begin{figure}[h!]
	{\centering
		\subfloat[]{\includegraphics[width=2.2in]{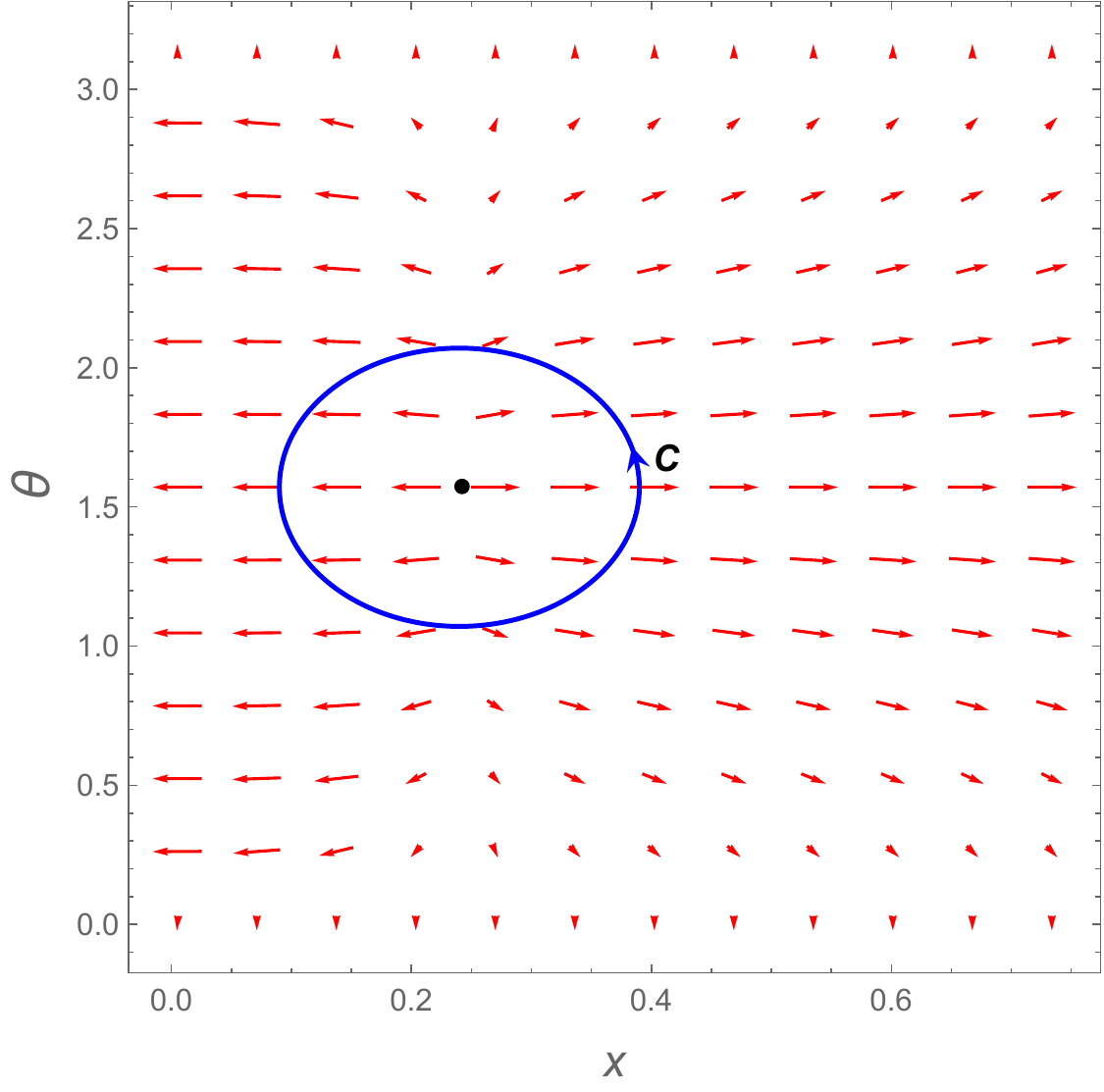}\label{Fig:bulk_vec_plot}}\hspace{0.8cm}	
		\subfloat[]{\includegraphics[width=3in]{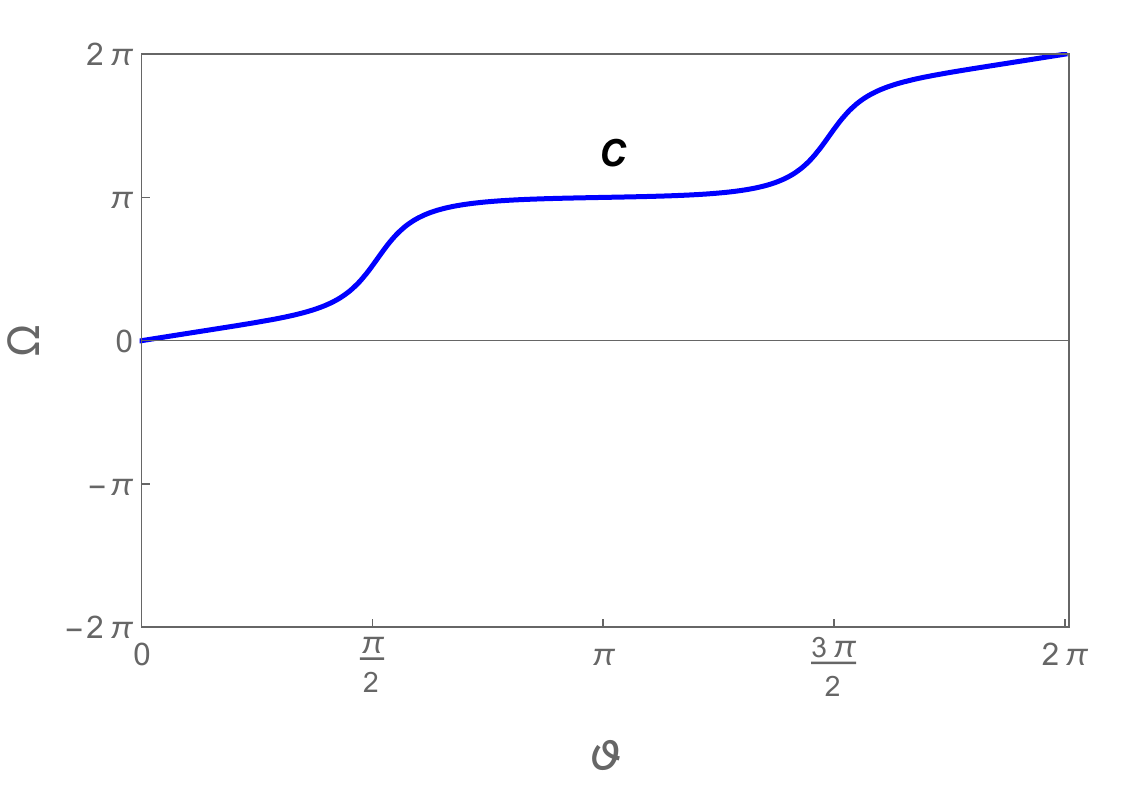}\label{Fig:bulk_omega_plot}}				
		
		\caption{\footnotesize (a) The normalised vector field  in the $\theta-x$ plane, shows the presence of Hawking-Page (HP) transition point (black dot, located at $x_{\rm HP} = 0.243$) at $\phi =0$. Contour $C$ encloses the HP transition point. 
			 (b) $\Omega$ vs $\vartheta$ for contour $C$. Plots are displayed for $n=4, \mu=0.2, \omega_3 =l=\beta=1$. } 
	}
\end{figure}

 In order to assign the topological charge to HP transition point, we employ the temperature $T_{\rm 0}^{\phantom{0}} (x, \mu)$ to define the vector filed $\phi(\phi^{x}, \phi^{\theta})$, in the following way:
\begin{eqnarray}
\phi^{x} &=& (\partial_{x} \Phi )_{\theta, \mu} \ \\
\phi^{\theta} &=& (\partial_{\theta} \Phi)_{x, \mu} \ 
\end{eqnarray}
where, $\Phi=\frac{1}{\text{sin}\theta} T_{\rm 0}^{\phantom{0}} (x, \mu)$.
This vector field $\phi$ vanishes exactly at the Hawking-Page transition point, which can be seen clearly from the normalised vector field  $(\frac{\phi^{x}}{||\phi||},\frac{\phi^\theta}{||\phi||})$  plot in the Fig.~\ref{Fig:bulk_vec_plot}.
The behaviour of the deflection angle $\Omega(\vartheta)$, for the given contour $C$ in the Fig.~\ref{Fig:bulk_vec_plot}, is as shown in Fig~\ref{Fig:bulk_omega_plot}, from where we find that the topological charge associated with the HP transition point would be $Q_t\big|_{\rm HP}^{\phantom{HP}} = \frac{1}{2\pi} \Omega (2\pi) = +1$.

\section{Boundary gauge theory}\label{3}
\begin{figure}[h!]
 	
 	{\centering
 		
 		\subfloat[]{\includegraphics[width=2.8in]{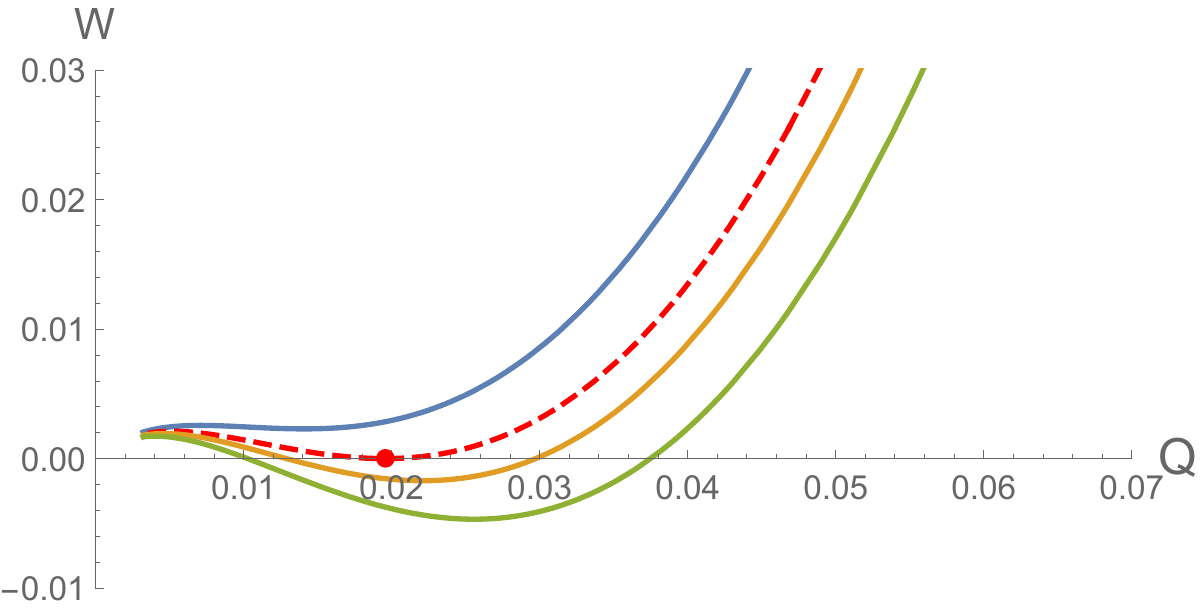}\label{fig:boundary_W}}\hspace{0.75cm}	
 		\subfloat[]{\includegraphics[width=2.6in]{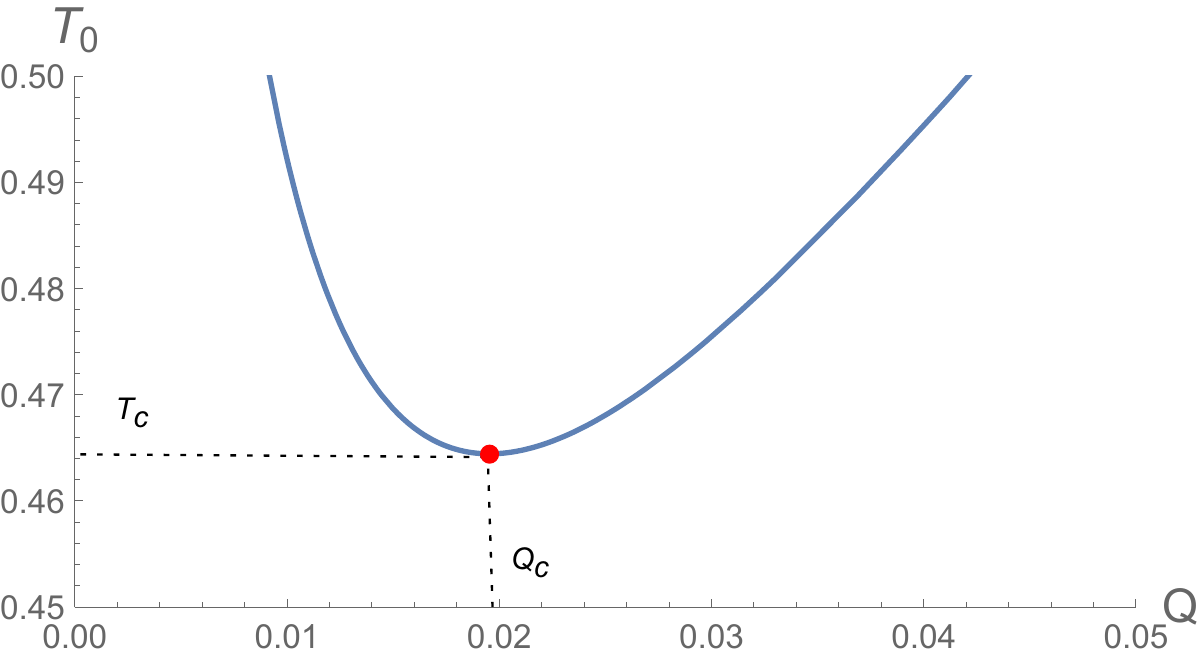}\label{fig:boundary_T0}}	
 		
 		\caption{\footnotesize For boundary gauge theory: (a) The effective potential  $W$ is plotted against $Q$ (using $x$ as a parameter)  at different temperatures $T$ for a fixed $\mu$. Temperature of the curves increases from top to bottom. Confinement-deconfinement transition happens for  the temperature $T_{\rm c}$ (dashed red curve) at  $Q_{\rm c}$ (red dot).  (b) The  curve $T_{\rm 0}$, as a function of $Q$ (using $x$ as a parameter), shows the confinement-deconfinement transition point at its minima (red dot).  Here, $(T_{\rm c}, Q_{\rm c}, x_{\rm c})= (0.464, 0.0196, 0.243)$. Plots are displayed for $n=4, \mu=0.2, N_c=\omega_3 =l=\beta=1$.}
 	}
 	
 \end{figure}

The effective potential in the gauge theory dual to Born-Infeld AdS black holes (in grand canonical ensemble) in $n=4$ is~\cite{Banerjee:2010ng}:
\begin{eqnarray}
W &=& \omega_3\,N_c^2 \bigg\{ -\frac{Tr_+^3}{2\pi} -Q\mu+\frac{3}{8\pi^2}\bigg( \frac{r_+^4}{l^2} + r_+^2 + \frac{\beta^2r_+^4}{3}\Big(1-\sqrt{1+\frac{4\pi^4 Q^2}{\beta^2 r_+^6}}\Big) \nonumber \\ 
&& + \frac{2\pi^4 Q^2 \,  _2F_1\big[\frac{1}{3}, \frac{1}{2}, \frac{4}{3}, -\frac{4\pi^4Q^2}{\beta^2 r_+^6}\big]}{r_+^2}\bigg) \bigg\}.
\end{eqnarray}
 where, $N_c$ stands for number of colours. In order to study the phase structure, again we use the parameter $x$ defined in eqn.~\eqref{eq:x}, and carry out
 a parametric plot of $W$ against $Q$, the order parameter in the boundary theory. The resulting phase structure [Fig.~\ref{fig:boundary_W}] shows a confinement-deconfinement transition at  critical temperature $T_c$, which is exactly at HP temperature $T_{\rm HP}$.
 \\ \noindent 
The critical temperature for the confinement-deconfinement transition can be obtained on satisfying the following two conditions simultaneously:
\begin{equation}\label{eq:cd_condition}
W = 0, \, \text{and}, \, \frac{\partial W}{\partial Q} = 0.
\end{equation}
Alternatively, as shown in the Fig.~\ref{fig:boundary_T0}, one can also find $T_c$ from the minima of the curve $T_0$, using  $W=0$. 
We note here that, using the parameter $x$, the confinement-deconfinement point can again be located at the  minima of the curve $T_0(x, \mu)$.
\begin{figure}[h!]
	{\centering
		\subfloat[]{\includegraphics[width=2.2in]{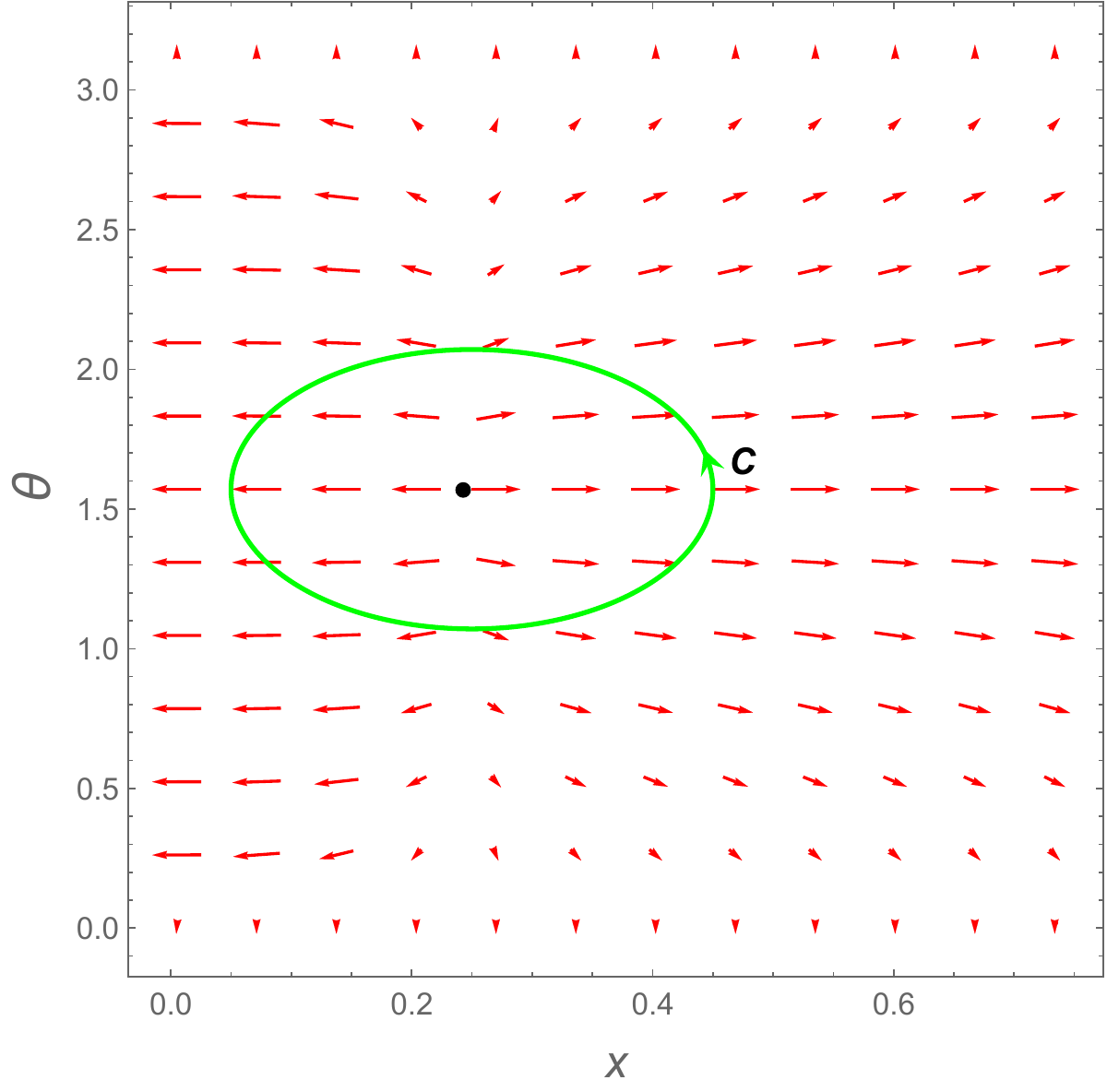}\label{Fig:boundary_vec_plot}}\hspace{0.75cm}	
		\subfloat[]{\includegraphics[width=3in]{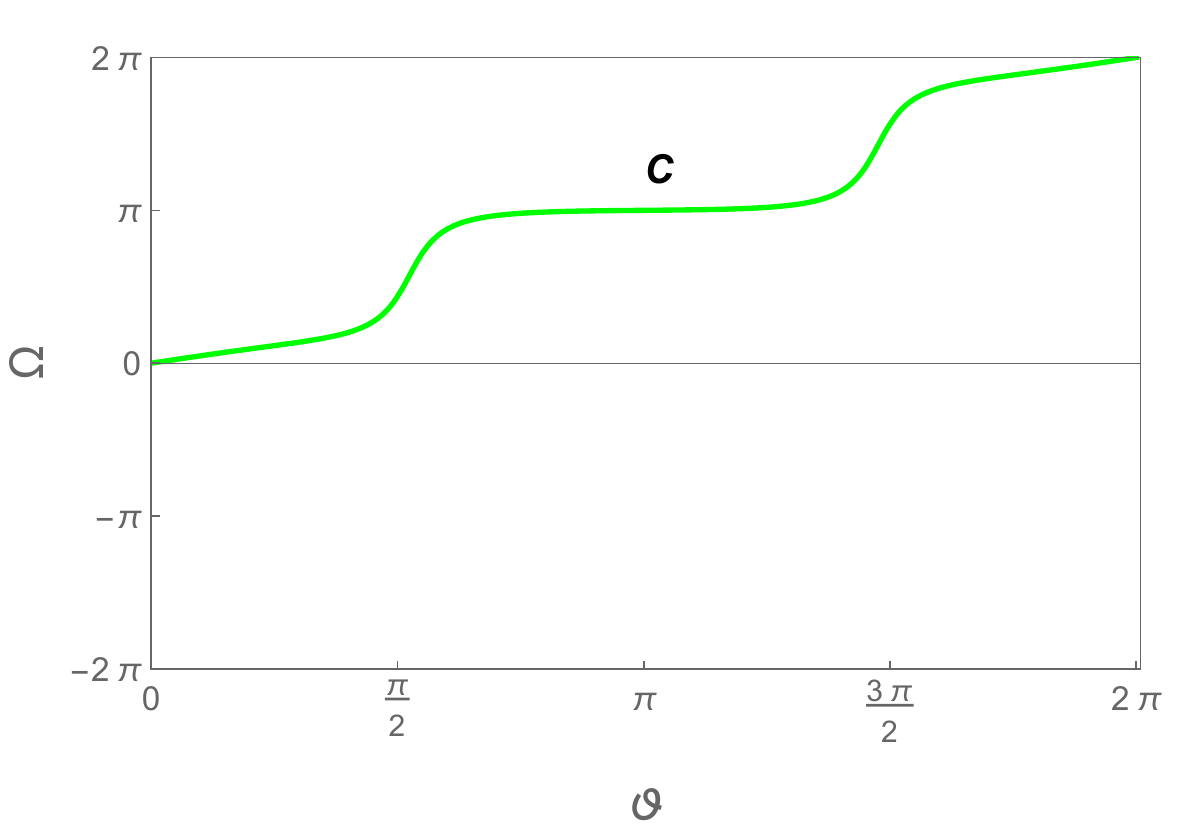}\label{Fig:boundary_omega_plot}}				
		
		\caption{\footnotesize For boundary gauge theory: (a) The  vector field $\phi$ vanishes at the confinement-deconfinement transition point (black dot, located at $x_{\rm c}$). Contour $C$ encloses the  transition point. 
			(b) $\Omega$ vs $\vartheta$ for contour $C$. Plots are displayed for $n=4, \mu =0.2, N_c=\omega_3=l=\beta=1$.} 
	}
\end{figure}
\noindent
Now, using $T_0(x, \mu)$, we define the vector filed $\phi(\phi^{x}, \phi^{\theta})$, as:
\begin{eqnarray}
\phi^{x} &=& (\partial_{x} \Phi )_{\theta, \mu} \ \\
\phi^{\theta} &=& (\partial_{\theta} \Phi)_{x, \mu} \ 
\end{eqnarray}
where, $\Phi=\frac{1}{\text{sin}\theta} T_0 (x, \mu)$.
This vector field $\phi$ vanishes exactly at the confinement-deconfinement transition point, as can be seen from the Fig.~\ref{Fig:boundary_vec_plot}. Then, the topological charge corresponding to this transition point turns out to be  $Q_t\big|_{\rm c}^{\phantom{c}} = \frac{1}{2\pi} \Omega (2\pi) = +1$ (given by the contour $C$ in Fig.~\ref{Fig:boundary_omega_plot}). This charge exactly matches the one obtained from bulk calculation.

\section{Conclusions}\label{4}
Following the topological classification of phase transition points of black holes discussed in~\cite{Wei:2021vdx,Yerra:2022alz,Ahmed:2022kyv,Yerra:2022eov,Wei:2022dzw,Yerra:2022coh}, we computed the topological charge of the Hawking-Page transition point in Einstein-Born-Infeld black holes in AdS using an off shell Bragg-Williams free energy approach and found it to be $+1$. It was shown that the same value appears from an analogous computation performed using an effective potential in the boundary gauge theory. Topological ideas from other directions are being advanced to understand critical points with interesting connections to quantum error correction~\cite{Almheiri:2014lwa,Bao:2022agm}. It seems plausible to attempt a topological classification of critical points in matrix models and possibly general thermodynamic systems~\cite{Yerra:2023hui}. It is expected that classification will reveal important information about the nature of phase transitions in a wide range of theories.

\ack
C.B. thanks IIT Bhubaneswar for financial support, and the DST (SERB), Government of India, through the Mathematical Research Impact Centric Support (MATRICS) grant no. MTR/2020/000135.
C.B. also thanks Renann Lipinski Jusinskas, the Institute of Physics (FZU) of the Czech Academy of Sciences \& CEICO, Prague, for warm hospitality, and the organisers of the XII. International Symposium on Quantum Theory and Symmetries (QTS12), Czech Technical University, Prague for a nice conference.
 
\appendix

\bibliographystyle{iopart-num}
\bibliography{iopart-num}

\end{document}